\documentclass[pra,aps,onecolumn,showpacs,superscriptaddress]{revtex4}
\usepackage{epsfig}
\usepackage{float}
\usepackage{latexsym,amssymb,amsthm,amsmath}

\begin{document}

\title{Three-qubit Grover's algorithm using superconducting quantum
interference devices in cavity-QED}
\author{Muhammad Waseem}
\affiliation{Department of Physics and Applied Mathematics, Pakistan Institute of
Engineering and Applied Sciences, Nilore, Islamabad, Pakistan.}
\author{Rizwan Ahmed}
\affiliation{Department of Physics and Applied Mathematics, Pakistan Institute of
Engineering and Applied Sciences, Nilore, Islamabad, Pakistan.}
\affiliation{Photonics Division, National Institute of Lasers and Optronics, Islamabad,
Pakistan.}
\author{Muhammad Irfan}
\email{m.irfanphy@gmail.com}
\affiliation{Department of Physics and Applied Mathematics, Pakistan Institute of
Engineering and Applied Sciences, Nilore, Islamabad, Pakistan.}
\author{Shahid Qamar}
\affiliation{Department of Physics and Applied Mathematics, Pakistan Institute of
Engineering and Applied Sciences, Nilore, Islamabad, Pakistan.}
\date{\today}

\begin{abstract}
We present a scheme for the implementation of three-qubit Grover's algorithm
using four-level superconducting quantum interference devices (SQUIDs)
coupled to a superconducting resonator. The scheme is based on resonant,
off-resonant interaction of the cavity field with SQUIDs and application of
classical microwave pulses. We show that adjustment of SQUID level spacings
during the gate operations, adiabatic passage, and second-order detuning are
not required that leads to faster implementation. We also show that the
marked state can be searched with high fidelity even in the presence of
unwanted off-resonant interactions, level decay, and cavity dissipation. 
\newline

\end{abstract}

\maketitle

\section{INTRODUCTION}

The quantum algorithms work much more efficiently than their classical\
counter parts due to quantum superposition and quantum interference. For
example, consider the search of an item in an unsorted database containing $%
N $ elements. Classical computation requires $O(N)$ steps to carry out the
search. However, Grover showed that search can be carried out with only $O(%
\sqrt{N})$ steps \cite{grover, farhi}. Thus, Grover's algorithm represents a
quadratic advantage over its classical counterpart.

Grover's algorithm has been realized using many physical systems like NMR 
\cite{iln}, superconducting qubits \cite{lyan, phyc} and atom cavity QED
systems \cite{pnas,harg,deng,wlyang,joshi}. Superconducting qubit cavity QED
is an attractive approach for quantum information processing due to their
strong coupling limit in microwave cavity as compared to atoms in cavity QED 
\cite{cpyang,cpyang2, wall}. SQUIDs have attracted much attention among the
superconducting qubits, due to their design flexibility, large-scale
integration, and compatibility to conventional electronics \textbf{\cite%
{hanx, ich, mooj}}. Recently, DiCarlo et al. demonstrated the implementation
of two-qubit Grover and Deutsch-Jozsa algorithms \cite{Carlo} and
preparation and measurement of three-qubit entanglement \cite{reed} using
superconducting qubits.

The goal of this work is to implement three-qubit Grover's algorithm using
four-level SQUIDs in cavity-QED. We consider a three-qubit phase gate, that
reduces the number of quantum gates typically required for the realization
of Grover's algorithm. Three-qubit Grover's algorithm is probabilistic \cite%
{wlyang}, as compared to two-qubit Grover's algorithm. Therefore, to achieve
high success probability, we have to implement basic searching iteration
several times. Implementation of three-qubit Grover search is much more
complex as compared to two-qubit case. In our scheme, two lowest energy
levels $\left\vert 0\right\rangle $ and $\left\vert 1\right\rangle $ of each
SQUID represent logical states. The scheme is based on resonant,
off-resonant interaction of cavity field with $\left\vert 2\right\rangle
\rightarrow \left\vert 3\right\rangle $ transition of SQUID and application
of resonant microwave pulses.

Our scheme does not require adjustment of SQUID level spacing during the
implementation of Grover's search iteration, thus, decoherence caused by the
adjustment of level spacing is suppressed. We do not require identical
coupling constants of each SQUID with the resonator and direct coupling
between the levels $\left\vert 1\right\rangle $ and $\left\vert
0\right\rangle $ \cite{lap}. Grover's iteration time becomes faster due to
resonant and off-resonant interactions as compared to second order detuning
or adiabatic passage.

Grover's iterations based on three-qubit quantum phase gate employed here,
considerably simplify the implementation as compared to conventional gate
decomposition method \cite{nielsen}. More importantly, it reduces the
possibility of error in comparison with a series of two-qubit gates. We also
consider the effect of spontaneous decay rate from intermediate level $%
\left\vert 3\right\rangle $ and decay of cavity field during the
implementation of Grover's iterations.

\section{GROVER'S ALGORITHM}

The basic idea of Grover's algorithm is as follows; we prepare input basis
states in superposition state $\left\vert \psi \right\rangle =(1/\sqrt{N}%
)\sum\nolimits_{i=0}^{N-1}\left\vert i\right\rangle $ by applying
Walsh-Hadamard transformation. First we, invert phase of the desired basis
state through unitary operator (called Oracle) and then invert all the basis
states about the average. We consider the implementation of Grover's
algorithm in terms of quantum logic networks as shown in Fig. 1. Any quantum
logical network can be constructed using quantum phase gates and
single-qubit quantum gates. The single-qubit quantum gate for $jth$ qubit
can be written in Dirac notation as 
\begin{equation}
H_{\theta ,\varphi }^{(j)}=\cos \theta I-i\sin \theta (e^{-i\varphi
}\left\vert 0\right\rangle \left\langle 1\right\vert +e^{i\varphi
}\left\vert 1\right\rangle \left\langle 0\right\vert ).  \label{eq1}
\end{equation}%
For $\theta =\pi /2$ and $\varphi =0$, we have $H_{2}^{(j)}=-i\sigma
_{x_{j}} $. Here $\sigma_{x}$ is the Pauli rotation matrix whose function is
to flip the state of qubit such that $\left\vert
0\right\rangle\rightarrow\left\vert 1\right\rangle $ and $\left\vert
1\right\rangle\rightarrow\left\vert 0\right\rangle .$ For $\theta =\pi /4$
and $\varphi =-\pi /2$, we have $H_{1}^{(j)}=H_{\pi /4,-\pi /2}^{j}$ \ which
transforms each qubit into superposition state i.e., $\left\vert
0\right\rangle \rightarrow (\left\vert 0\right\rangle +\left\vert
1\right\rangle )/\sqrt{2}$ and $\left\vert 1\right\rangle \rightarrow $ $%
(\left\vert 1\right\rangle -\left\vert 0\right\rangle )/\sqrt{2}.$

The transformation for three-qubit quantum controlled phase gate can be
expressed by 
\begin{equation}
Q_{\varPhi }\left\vert q_{1},q_{2},q_{3}\right\rangle =e^{(i\varPhi \delta
_{q_{1},1}\delta _{q_{2},1}\delta _{q_{3},1})}\left\vert
q_{1},q_{2},q_{3}\right\rangle ,  \label{eq2}
\end{equation}%
where $\left\vert q\right\rangle _{1}$, $\left\vert q\right\rangle _{2}$,
and $\left\vert q\right\rangle _{3}$ stand for basis $\left\vert
0\right\rangle $ or $\left\vert 1\right\rangle $ of the qubit and $\delta
_{q_{1},1}$, $\delta _{q_{2},1}$, and $\delta _{q_{3},1}$ are the Kroneker
delta functions. Thus, three-qubit quantum phase gate induces a phase $e^{i%
\varPhi }$ only when all three input qubit are in state $\left\vert
1\right\rangle $. Three-qubit quantum phase gate operator for $\varPhi =\pi $
can be written in Dirac notation as

\begin{eqnarray}
Q_{\pi } &=&\left\vert 000\right\rangle \left\langle 000\right\vert
+\left\vert 001\right\rangle \left\langle 001\right\vert +\left\vert
010\right\rangle \left\langle 010\right\vert  \notag \\
&&+\left\vert 011\right\rangle \left\langle 011\right\vert +\left\vert
100\right\rangle \left\langle 100\right\vert +\left\vert 101\right\rangle
\left\langle 101\right\vert  \notag \\
&&+\left\vert 110\right\rangle \left\langle 110\right\vert -\left\vert
111\right\rangle \left\langle 111\right\vert .  \label{eq3}
\end{eqnarray}%
\qquad \qquad\ 

The three-qubit controlled phase gate can be used instead of involving
series of two-qubit gates. This method not only simplifies the
implementation but also reduces the probability of error. Figure. \ref{fig1}
shows the circuit diagram of three-qubit Grover's algorithm based on
three-qubit phase gate and two-qubit gates \cite{zubairy, zubairy2}.
Consider that the initial state of three qubits is $\left\vert
000\right\rangle $. Grover's algorithm can be carried out using the
following three steps:

\begin{figure}[tbp]
\includegraphics[width=3.6 in]{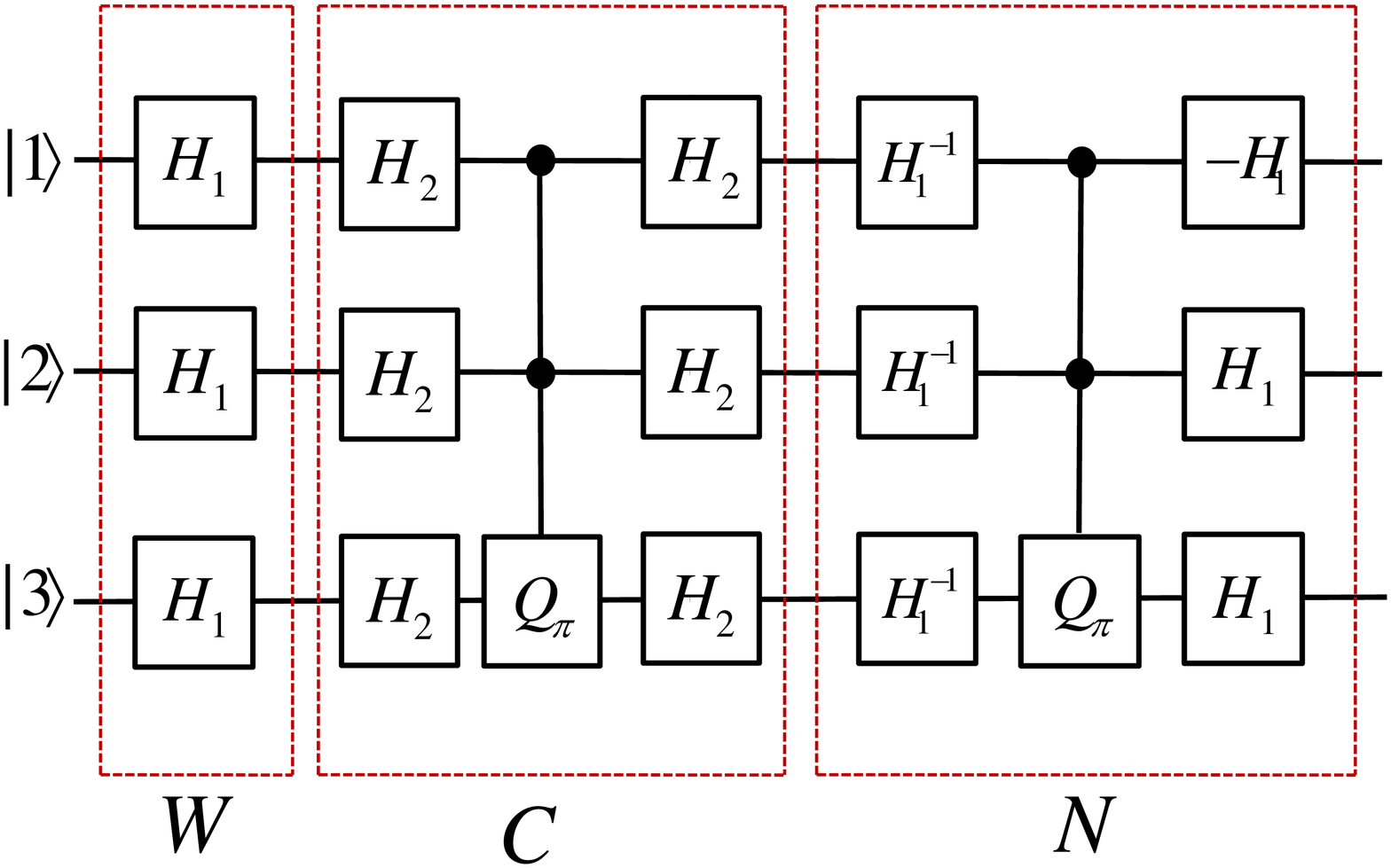}%
\caption{Circuit diagram for three-qubit Grover's algorithm.}
\label{fig1}
\end{figure}

\textit{Part 1}\textbf{\ (W)}: Apply Walsh-Hadamard transformation $%
W=H_{1}^{1}\otimes H_{1}^{2}\otimes H_{1}^{3}$ on each qubit. The resultant
state is therefore given by 
\begin{align}
\left\vert \psi \right\rangle & =\frac{1}{2^{3/2}}(\left\vert
000\right\rangle +\left\vert 001\right\rangle +\left\vert 010\right\rangle
+\left\vert 011\right\rangle  \notag \\
& +\left\vert 100\right\rangle +\left\vert 101\right\rangle +\left\vert
110\right\rangle +\left\vert 111\right\rangle ).  \label{eq4}
\end{align}

\textit{Part 2 }(\textbf{C}): In this step, consider the unitary operator $%
C=C_{q_{1},q_{2},q_{3}}$ (called Oracle) which changes the sign of target
state $\left\vert q_{1},q_{2},q_{3}\right\rangle $. The operator $%
I-2\left\vert q_{1},q_{2},q_{3}\right\rangle \left\langle
q_{1},q_{2},q_{3}\right\vert $ performs the unitary transformation which can
be implemented using three-qubit phase gate $Q_{\pi }$ and single-qubit gate 
$H_{2}=-i$ $\sigma _{x_{j}}$ as shown in Fig. \ref{fig1}. The sign change
operators for eight possible target states are given by 
\begin{eqnarray}
C_{000} &=&-\sigma _{x1}\sigma _{x2}\sigma _{x3}Q_{\pi }\sigma _{x1}\sigma
_{x2}\sigma _{x3},  \notag \\
\text{ }C_{111}&=& Q_{\pi },  \notag \\
\text{ }C_{001} &=&\sigma _{x1}\sigma _{x2}Q_{\pi }\sigma _{x1}\sigma _{x2},
\notag \\
C_{010}&=&\sigma _{x1}\sigma _{x3}U_{\pi }\sigma _{x1}\sigma _{x3},  \notag
\\
\text{ }C_{011} &=&-\sigma _{x1}Q_{\pi }\sigma _{x1},  \notag \\
C_{100}&=&\sigma_{x2}\sigma _{x3}Q_{\pi }\sigma _{x2}\sigma _{x3},  \notag \\
\text{ }C_{101} &=&-\sigma _{x2}Q_{\pi }\sigma _{x2},  \notag \\
C_{110}&=&-\sigma_{x3}Q_{\pi }\sigma _{x3}.  \label{equ5}
\end{eqnarray}%
Now Oracle applies one of $C_{q_{1},q_{2},q_{3}}$ operators on state given
in Eq. (\ref{eq4}) and changes the sign of target state. For example, our
target state is $\left\vert 001\right\rangle $, then by applying $C_{001}$
on state (\ref{eq4}), we obtain the change of phase on target state $%
\left\vert 001\right\rangle $ i.e., 
\begin{align}
\left\vert \psi _{1}\right\rangle & =\frac{1}{2^{3/2}}(\left\vert
000\right\rangle -\left\vert 001\right\rangle +\left\vert 010\right\rangle
+\left\vert 011\right\rangle  \notag \\
& +\left\vert 100\right\rangle +\left\vert 101\right\rangle +\left\vert
110\right\rangle +\left\vert 111\right\rangle ).  \label{eq6}
\end{align}

\textit{Part 3 }(\textbf{N}): In this step, our goal is to find out the
marked state $\left\vert 001\right\rangle $. This can be accomplished
through inversion about mean using the operator $N=I-2\left\vert \psi
\right\rangle _{1}\left\langle \psi \right\vert $. It is clear from Fig. \ref%
{fig1} that the operator $N$ can be written in terms of single-qubit gate
and three qubit quantum phase gate 
\begin{equation}
N=\{(-H_{1}^{1}\otimes H_{1}^{2}\otimes H_{1}^{3})\}Q_{\pi
}\{(H_{1}^{1})^{-1}\otimes (H_{1}^{2})^{-1}\otimes (H_{1}^{3})^{-1}\}.
\label{eq7}
\end{equation}%
The combined operator $G=-NC_{q_{1},q_{2},q_{3}}$ is called Grover's
operator.\ When $G$ is applied to initial state $\left\vert \psi
\right\rangle ,$ $k$ $(\thickapprox \pi \sqrt{N}/4)$ times, then the
probability of searching target state becomes maximum \cite{farhi}.

\section{Basic Theory}

Here, we consider rf-SQUIDs as qubits that consist of a single Josephson
junction enclosed by superconducting loop. The corresponding Hamiltonian is
given by \cite{rouse} 
\begin{equation}
H_{S}=\frac{Q_{c}^{2}}{2C}+\frac{(\phi -\phi _{x})^{2}}{2L}-E_{J}cos(\frac{%
2\pi \phi }{\phi _{0}}),  \label{eq8}
\end{equation}%
where $C$ and $L$ are junction capacitance and loop inductance,
respectively. The conjugate variables of the system are magnetic flux $\phi $
threading the ring and total charge $Q_{c}$ on capacitor. The static
external flux applied to the ring is $\phi _{x}$ and $E_{J}\equiv \frac{%
I_{c}\phi _{0}}{2\pi }$ is the Josephson coupling energy. Here, $I_{c}$ is
critical current of Josephson junction and $\phi _{0}=\frac{h}{2e}$ is the
flux quantum.

\begin{figure}[tbp]
\includegraphics[width=3.4 in]{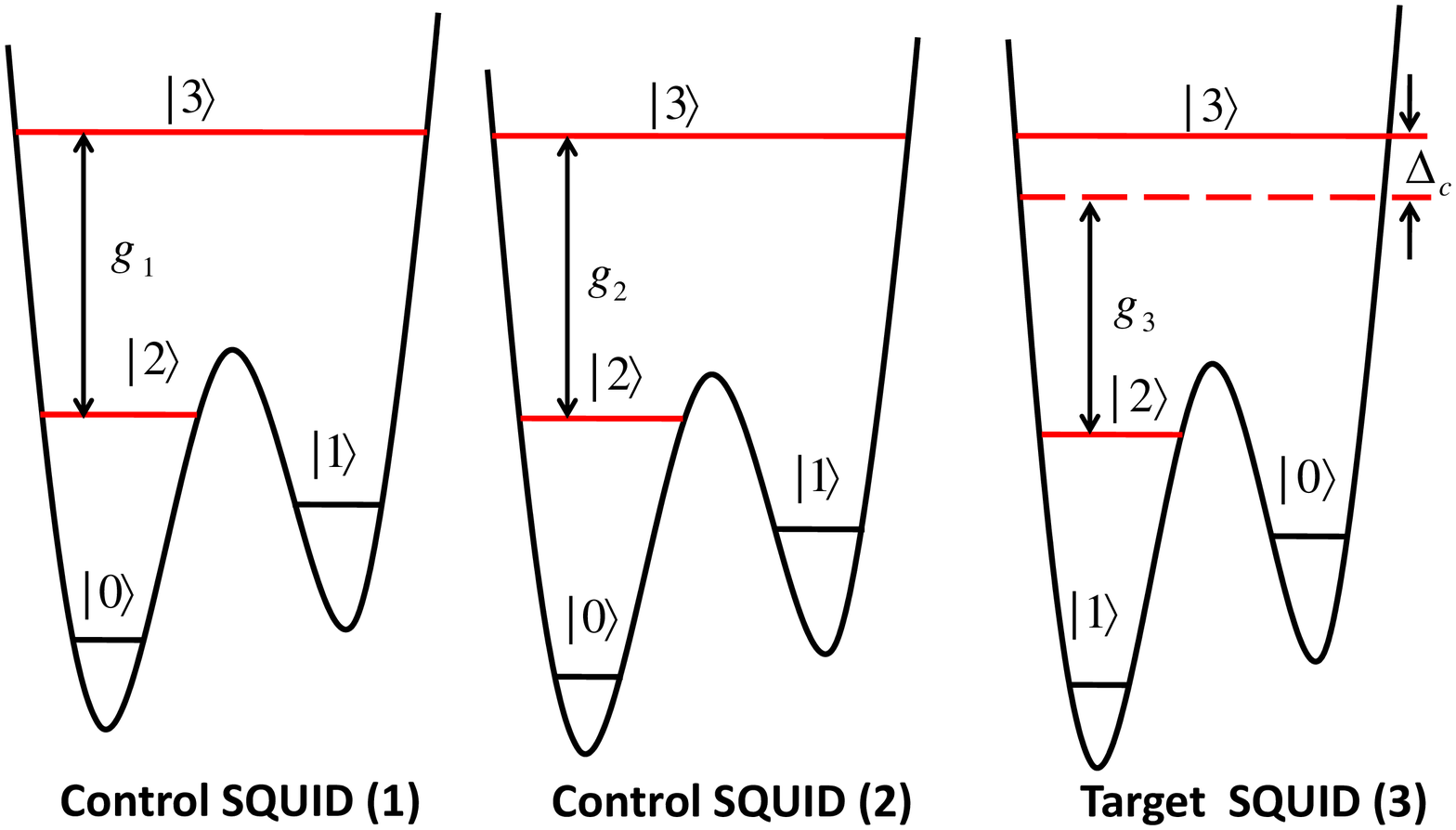}
\caption{Resonant interaction of SQUID (1) and (2) with cavity field and off
resonant interaction of target SQUID (3).}
\label{fig2}
\end{figure}

The SQUIDs are biased properly to achieve desired four-level structure as
shown in Fig. \ref{fig2} by varying the external flux \cite{w}. The
single-mode of the cavity field is resonant with $\left\vert 2\right\rangle
\leftrightarrow \left\vert 3\right\rangle $ transition of SQUIDs 1 and 2.
The evolution of initial state $\left\vert 3\right\rangle \left\vert
0\right\rangle _{c}$ and $\left\vert 2\right\rangle \left\vert
1\right\rangle _{c}$ under the effect of Hamiltonian (\ref{eq8}) can be
written as \cite{w} 
\begin{align}
\left\vert 3\right\rangle \left\vert 0\right\rangle _{c}& \rightarrow
cos(g_{i}t)\left\vert 3\right\rangle \left\vert 0\right\rangle
_{c}-isin(g_{i}t)\left\vert 2\right\rangle \left\vert 1\right\rangle _{c}, 
\notag \\
\left\vert 2\right\rangle \left\vert 1\right\rangle _{c}& \rightarrow
cos(g_{i}t)\left\vert 2\right\rangle \left\vert 1\right\rangle
_{c}-isin(g_{i}t)\left\vert 3\right\rangle \left\vert 0\right\rangle _{c},
\label{eq9}
\end{align}%
where $\left\vert 0\right\rangle _{c}$ ($\left\vert 1\right\rangle _{c}$) is
vacuum (single photon) state of the cavity field and $g_{i}$ ($i=1,2$) is
the coupling constant between the cavity field and $\left\vert
2\right\rangle \leftrightarrow \left\vert 3\right\rangle $ transition of the
SQUID $1$ and $2$.

The cavity field interacts off-resonantly with $\left\vert 2\right\rangle
\leftrightarrow \left\vert 3\right\rangle $ transition of SQUID 3 (i.e., $%
\Delta _{c}=\omega _{c}-\omega _{32}>>g_{3})$ as shown in Fig. \ref{fig2}.
Here, $\Delta _{c}$ is the detuning between $\left\vert 2\right\rangle
\leftrightarrow \left\vert 3\right\rangle $ transition frequency $\omega
_{32}$ of SQUID 3, $\omega _{c}$ is the frequency of resonator and $g_{3}$
is coupling constant between resonator mode and $\left\vert 2\right\rangle
\leftrightarrow \left\vert 3\right\rangle $ transition. In the presence of
single photon inside the cavity, the evolution of initial state $\left\vert
2\right\rangle \left\vert 1\right\rangle _{c}$ and $\left\vert
3\right\rangle \left\vert 1\right\rangle _{c}$ are given by \cite{w} 
\begin{eqnarray}
\left\vert 2\right\rangle \left\vert 1\right\rangle _{c} &\rightarrow
&e^{ig_{3}^{2}t/\Delta _{c}}\left\vert 2\right\rangle \left\vert
1\right\rangle _{c},  \notag \\
\left\vert 3\right\rangle \left\vert 1\right\rangle _{c} &\rightarrow
&e^{-ig_{3}^{2}t/\Delta _{c}}\left\vert 3\right\rangle \left\vert
1\right\rangle _{c}.  \label{eq10}
\end{eqnarray}%
It is clear that phase shifts $e^{i\frac{g_{3}^{2}t}{\Delta _{c}}}$ and $%
e^{-i\frac{g_{3}^{2}t}{\Delta _{c}}}$ are induced to states $\left\vert
2\right\rangle $ and $\left\vert 3\right\rangle $ of SQUID 3, respectively.
However, states $\left\vert 2\right\rangle \left\vert 0\right\rangle _{c}$
and $\left\vert 3\right\rangle \left\vert 0\right\rangle _{c}$ remain
unchanged.

A classical microwave pulse resonant with $\left\vert i\right\rangle
\leftrightarrow \left\vert j\right\rangle \ $ ($i,j=0,1,2,3$) is applied to
each SQUID. The evolution of states can be written as \cite{yang} 
\begin{align}
\left\vert i\right\rangle & \rightarrow cos(\Omega _{ij}t)\left\vert
i\right\rangle -ie^{-i\varphi }sin(\Omega _{ij}t)\left\vert j\right\rangle ,
\notag \\
\left\vert j\right\rangle & \rightarrow cos(\Omega _{ij}t)\left\vert
j\right\rangle -ie^{i\varphi }sin(\Omega _{ij}t)\left\vert i\right\rangle ,
\label{eq11}
\end{align}%
where $\Omega _{ij}$ is the Rabi frequency between two levels $\left\vert
i\right\rangle $ and $\left\vert j\right\rangle $ and $\varphi $ is the
phase associated with classical field. It may be mentioned that resonant
interaction between pulse and SQUID can be carried out in a very short time
by increasing the Rabi frequency of pulse.

\section{Implementation of Grover's algorithm}

Three SQUIDs shown in Fig. \ref{fig2} are initially prepared in state $%
\left\vert 000\right\rangle $. \ For notation convenience, we denote ground
level as $\left\vert 1\right\rangle $ and first excited state as $\left\vert
0\right\rangle $ for SQUID $3$ as shown in Fig. \ref{fig2}.

\textit{Part 1}\textbf{\ (W)}:\textit{\ }To accomplish \textit{part 1}of
Grover's algorithm, we apply single-qubit gate $\ H_{1}^{j}$ to each SQUID
as shown in Fig. \ref{fig1}. The single-qubit gate $H_{1}^{j}$ is carried
out through two-step process that involves an auxiliary level $\left\vert
3\right\rangle $ via method described in Ref \cite{yh}. We need two
microwave pulses of different frequencies resonant to $\left\vert
1\right\rangle \leftrightarrow \left\vert 3\right\rangle $ and $\left\vert
0\right\rangle \leftrightarrow \left\vert 3\right\rangle $ transitions. The
desired arbitrary single-qubit gate (See Eq. (1)) can be achieved by
choosing a proper interaction time (i.e., $\theta =\Omega _{03}t$ ) and
phase $\varphi $ of classical microwave pulse resonance to $\left\vert
0\right\rangle \leftrightarrow \left\vert 3\right\rangle $ transition. We
achieve $H_{1}^{j}$ by choosing $\Omega _{03}t=\pi /4$ and $\varphi =-\pi
/4, $ as a result we obtain state given by Eq. (\ref{eq4}).

\textit{Part 2 }(\textbf{C})\textbf{:} In order to implement $%
C=C_{q_{1},q_{2},q_{3}},$ apply single-qubit gate $H_{2}=-i$ $\sigma
_{x_{j}} $ by choosing $\Omega _{03}t=\pi /2$ and $\varphi =0$ as shown in
Fig. \ref{fig1}. Then apply three-qubit quantum controlled phase gate. The
procedure for realizing the three-qubit controlled phase gate is described
as follows:

Initially, cavity is in a vacuum state $\left\vert 0\right\rangle _{c}$ and
levels $\left\vert 2\right\rangle $ and $\left\vert 3\right\rangle $ of each
SQUID are not occupied.

\textsl{Step 1.} Apply microwave pulse (with $\Omega _{13}t_{1}=\frac{\pi }{2%
}$ and phase $\varphi =\pi $, where $t_{1}$ is pulse duration) resonant to $%
\left\vert 1\right\rangle \leftrightarrow \left\vert 3\right\rangle $
transition of the SQUID $1$ to occupy level $\left\vert 3\right\rangle _{1}.$
Cavity field interacts resonantly to the $\left\vert 2\right\rangle
_{1}\leftrightarrow \left\vert 3\right\rangle _{1}$ transition of SQUID $1$
for time interval $t_{1}^{^{\prime }}=\pi /2g_{1}$ such that the
transformation $\left\vert 3\right\rangle _{1}\left\vert 0\right\rangle
_{c}\rightarrow -i\left\vert 2\right\rangle _{1}\left\vert 1\right\rangle
_{c}$ is obtained. The overall step can be written as $\left\vert
1\right\rangle _{1}\left\vert 0\right\rangle _{c}\rightarrow \left\vert
2\right\rangle _{1}\left\vert 1\right\rangle _{c}$. However, the state $%
\left\vert 0\right\rangle _{1}\left\vert 0\right\rangle _{c}$ remains
unchanged.

\textsl{Step 2.} Apply microwave pulse (with $\Omega _{20}t_{2}=\pi /2$ and
phase $\varphi =\pi /2$) to the SQUID 1 while a microwave pulse (with $%
\Omega _{20}t_{2}=\pi /2$ and phase $\varphi =-\pi /2$) to the SQUID 2. As a
result transformations $\left\vert 2\right\rangle _{1}(\left\vert
0\right\rangle _{1})\rightarrow \left\vert 0\right\rangle _{1}(-\left\vert
2\right\rangle _{1})$ for SQUID 1 while $\left\vert 0\right\rangle
_{2}(\left\vert 2\right\rangle _{2})\rightarrow \left\vert 2\right\rangle
_{2}(-\left\vert 0\right\rangle _{2})$ for SQUID 2 are obtained.

\textsl{Step 3.} After the above operation, when cavity is in a single
photon state $\left\vert 1\right\rangle_{c} $, only the level $\left\vert
2\right\rangle $ of SQUID 2 is populated. The cavity field interacts
resonantly to $\left\vert 2\right\rangle _{2}\rightarrow \left\vert
3\right\rangle _{2}$ transition of SQUID 2 for time $t_{3}^{^{\prime }}=\pi
/2g_{2}.$ We then obtain transformation $\left\vert 2\right\rangle
_{2}\left\vert 1\right\rangle _{c}\rightarrow -i\left\vert 3\right\rangle
_{2}\left\vert 0\right\rangle _{c},$ while states $\left\vert 2\right\rangle
_{2}\left\vert 0\right\rangle _{c}$, $\left\vert 1\right\rangle
_{2}\left\vert 0\right\rangle _{c}$ and, $\left\vert 1\right\rangle
_{2}\left\vert 1\right\rangle _{c}$ remain unchanged. Then apply microwave
pulse (with $\Omega _{30}t_{3}=\pi /2$ and phase $\varphi =\pi $) to SQUID 2
to transform state $\left\vert 3\right\rangle _{2}$ to $i\left\vert
0\right\rangle _{2}$. The overall step can be written as $\left\vert
2\right\rangle _{2}\left\vert 1\right\rangle _{c}\rightarrow \left\vert
0\right\rangle _{2}\left\vert 0\right\rangle _{c}$. However, states $%
\left\vert 1\right\rangle _{2}\left\vert 0\right\rangle _{c}$, $\left\vert
1\right\rangle _{2}\left\vert 1\right\rangle _{c}$ and $\left\vert
2\right\rangle _{2}\left\vert 0\right\rangle _{c}$ remain unchanged. The
evolution of the system after above three steps is given by

\begin{equation}
\begin{array}{c}
\left\vert 100\right\rangle \left\vert 0\right\rangle _{c} \\ 
\left\vert 101\right\rangle \left\vert 0\right\rangle _{c} \\ 
\left\vert 110\right\rangle \left\vert 0\right\rangle _{c} \\ 
\left\vert 111\right\rangle \left\vert 0\right\rangle _{c}%
\end{array}%
\overset{1}{\rightarrow }%
\begin{array}{c}
\left\vert 200\right\rangle \left\vert 1\right\rangle _{c} \\ 
\left\vert 201\right\rangle \left\vert 1\right\rangle _{c} \\ 
\left\vert 210\right\rangle \left\vert 1\right\rangle _{c} \\ 
\left\vert 211\right\rangle \left\vert 1\right\rangle _{c}%
\end{array}%
\overset{2}{\rightarrow }%
\begin{array}{c}
\left\vert 020\right\rangle \left\vert 1\right\rangle _{c} \\ 
\left\vert 021\right\rangle \left\vert 1\right\rangle _{c} \\ 
\left\vert 010\right\rangle \left\vert 1\right\rangle _{c} \\ 
\left\vert 011\right\rangle \left\vert 1\right\rangle _{c}%
\end{array}%
\overset{3}{\rightarrow }%
\begin{array}{c}
\left\vert 000\right\rangle \left\vert 0\right\rangle _{c} \\ 
\left\vert 001\right\rangle \left\vert 0\right\rangle _{c} \\ 
\left\vert 010\right\rangle \left\vert 1\right\rangle _{c} \\ 
\left\vert 011\right\rangle \left\vert 1\right\rangle _{c}.%
\end{array}%
\end{equation}

It must be noted here that we have shown the evolution of only four initial
possible states out of eight states (See Eq. (4)) since the evolution of
other four states i.e., $\left\vert 000\right\rangle \left\vert
0\right\rangle _{c}, \left\vert 001\right\rangle \left\vert 0\right\rangle
_{c}, \left\vert 010\right\rangle \left\vert 0\right\rangle _{c},$ and $%
\left\vert 011\right\rangle \left\vert 0\right\rangle _{c}$ is trivial.

\textsl{Step 4.} Apply microwave pulse with $\Omega _{12}t_{4}=\pi /2$ and
phase $\varphi =-\pi /2$ to SQUID 3 to obtain transformation $\left\vert
1\right\rangle _{3}\rightarrow $ $\left\vert 2\right\rangle _{3}$. After the
above operation only level $\left\vert 2\right\rangle $ of SQUID (3) is
populated, when cavity is in a single photon state. Now the cavity field
interacts off-resonantly to $\left\vert 2\right\rangle _{3}\rightarrow
\left\vert 3\right\rangle _{3}$ transition of SQUID 3. It is clear from Eq. (%
\ref{eq10}) that for $t_{4}^{^{\prime }}=(\pi \Delta _{c})/g_{3}^{2}$, state 
$\left\vert 2\right\rangle _{3}\left\vert 1\right\rangle _{c}$ changes to $%
-\left\vert 2\right\rangle _{3}\left\vert 1\right\rangle _{c}$. However,
states $\left\vert 0\right\rangle _{3}\left\vert 0\right\rangle _{c}$, $%
\left\vert 0\right\rangle _{3}\left\vert 1\right\rangle _{c}$ and $%
\left\vert 2\right\rangle _{3}\left\vert 0\right\rangle _{c}$ remain
unchanged.

\textsl{Step 5.} Apply microwave pulse (with $\Omega _{12}t_{4}=\pi /2$ and
phase $\varphi =\pi /2$) to SQUID 3, as a result state transformation $%
\left\vert 2\right\rangle _{3}\rightarrow \left\vert 1\right\rangle _{3}$ is
obtained.

\textsl{Step 6.} Perform reverse of the operations mentioned in step 3.
Apply microwave pulse (with $\Omega _{30}t_{3}=\pi /2$ and phase $\varphi
=\pi $ ) to SQUID 2. Wait for time $t_{3}^{^{\prime }}$ given in step 3,
during which cavity field interacts resonantly to $\left\vert 2\right\rangle
_{2}\rightarrow \left\vert 3\right\rangle _{2}$ transition of the SQUID 2.
Over all transformation can easily be written as $\left\vert 0\right\rangle
_{2}\left\vert 0\right\rangle _{c}\rightarrow \left\vert 2\right\rangle
_{2}\left\vert 1\right\rangle _{c}$. However, states $\left\vert
1\right\rangle _{2}\left\vert 0\right\rangle _{c}$, $\left\vert
1\right\rangle _{2}\left\vert 1\right\rangle _{c}$ and $\left\vert
2\right\rangle _{2}\left\vert 0\right\rangle _{c}$ remain unchanged. The
evolution of the system after applying steps 4-6 is given by

\begin{equation}
\overset{4}{\rightarrow }%
\begin{array}{c}
\text{ \ }\left\vert 000\right\rangle \left\vert 0\right\rangle _{c} \\ 
\text{ \ }\left\vert 002\right\rangle \left\vert 0\right\rangle _{c} \\ 
\text{ \ }\left\vert 010\right\rangle \left\vert 1\right\rangle _{c} \\ 
-\left\vert 012\right\rangle \left\vert 1\right\rangle _{c}%
\end{array}%
\overset{5}{\rightarrow }%
\begin{array}{c}
\text{ \ }\left\vert 000\right\rangle \left\vert 0\right\rangle _{c} \\ 
\text{ \ }\left\vert 001\right\rangle \left\vert 0\right\rangle _{c} \\ 
\text{ \ }\left\vert 010\right\rangle \left\vert 1\right\rangle _{c} \\ 
-\left\vert 011\right\rangle \left\vert 1\right\rangle _{c}%
\end{array}%
\overset{6}{\rightarrow }%
\begin{array}{c}
\text{ \ \ }\left\vert 020\right\rangle \left\vert 1\right\rangle _{c} \\ 
\text{ \ \ }\left\vert 021\right\rangle \left\vert 1\right\rangle _{c} \\ 
\text{ \ \ }\left\vert 010\right\rangle \left\vert 1\right\rangle _{c} \\ 
-\left\vert 011\right\rangle \left\vert 1\right\rangle _{c}.%
\end{array}%
\end{equation}

\textsl{Step 7.} Apply microwave pulse (with $\Omega _{20}t_{2}=\pi /2$ and
phase $\varphi =\pi /2$) to SQUID 2 while a microwave pulse (with $\Omega
_{20}t_{2}=\pi /2$ and phase $\varphi =-\pi /2$) to SQUID 1. The
transformations $\left\vert 2\right\rangle _{2}(\left\vert 0\right\rangle
_{2})\rightarrow \left\vert 0\right\rangle _{2}(-\left\vert 2\right\rangle
_{2})$ for SQUID 2 while $\left\vert 2\right\rangle _{1}(\left\vert
0\right\rangle _{1})\rightarrow -\left\vert 0\right\rangle _{1}(\left\vert
2\right\rangle _{1})$ for SQUID 1 are obtained.

\textsl{Step 8.} Now perform reverse operation of step 1. First wait for
time interval $t_{1}^{^{\prime }}=\pi /2g_{1}$ during which resonator
interacts resonantly to the $\left\vert 2\right\rangle _{1}\leftrightarrow
\left\vert 3\right\rangle _{1}$ transition of SQUID 1. Then apply microwave
pulse (with $\Omega _{31}t_{1}=\pi /2$ and phase $\varphi =\pi $) to SQUID
1. The overall step can easily be written as $\left\vert 2\right\rangle
_{1}\left\vert 1\right\rangle _{c}\rightarrow \left\vert 1\right\rangle
_{1}\left\vert 0\right\rangle _{c}$. However, state $\left\vert
0\right\rangle _{1}\left\vert 0\right\rangle _{c}$ remains unchanged. After
applying steps 7-8, the system evolves as

\begin{equation}
\overset{7}{\rightarrow }%
\begin{array}{c}
\text{ \ }\left\vert 200\right\rangle \left\vert 1\right\rangle _{c} \\ 
\text{ \ }\left\vert 201\right\rangle \left\vert 1\right\rangle _{c} \\ 
\text{ \ }\left\vert 210\right\rangle \left\vert 1\right\rangle _{c} \\ 
-\left\vert 211\right\rangle \left\vert 1\right\rangle _{c}%
\end{array}%
\overset{8}{\rightarrow }%
\begin{array}{c}
\text{ \ }\left\vert 100\right\rangle \left\vert 0\right\rangle _{c} \\ 
\text{ \ }\left\vert 101\right\rangle \left\vert 0\right\rangle _{c} \\ 
\text{ \ }\left\vert 110\right\rangle \left\vert 0\right\rangle _{c} \\ 
-\left\vert 111\right\rangle \left\vert 0\right\rangle _{c}.%
\end{array}%
\end{equation}

After the application of three-qubit phase gate, the state given by Eq. (\ref%
{eq4}) evolves to

\begin{align}
\left\vert \psi _{3}\right\rangle & =\frac{1}{2^{3/2}}(\left\vert
000\right\rangle +\left\vert 001\right\rangle +\left\vert 010\right\rangle
+\left\vert 011\right\rangle  \notag \\
& +\left\vert 100\right\rangle +\left\vert 101\right\rangle +\left\vert
110\right\rangle -\left\vert 111\right\rangle ).  \label{eq12}
\end{align}
Next apply single-qubit gate i.e., $H_{2}=-i$ $\sigma _{x_{j}},$ which
completes the part 2 (C) of the implementation scheme.

\textit{Part 3 }(\textbf{N}): In order to implement operator $N$, apply
single-qubit gate $H_{1}^{-1}$, then apply three-qubit quantum controlled
phase gate by repeating the above mentioned 8 steps. Finally, apply
single-qubit Hadamard gate $H_{1}$. It is clear that Grover's operator $%
G=-NC_{q_{1},q_{2},q_{3}}$ for eight objects ($N=8 $) can be implemented
using four-level SQUIDs coupled to superconducting resonator.

It may be pointed out that in order to implement a three-qubit phase gate
using conventional decomposition method, it requires twenty five basic
gates, i.e., six two-qubit phase gates, twelve single-qubit Hadamard gates,
and seven single-qubit phase shift gates \cite{nielsen}. If we assume that
the realization of each basic gate requires a one-step operation only, then
twenty five steps are required for a three-qubit phase gate. Whereas in our
scheme total number of eight steps are needed to implement three-qubit phase
gate. As a result our proposed implementation scheme for Grover's search
algorithm which is based on three-qubit phase gate is faster than one based
on two-qubit phase gate.

After performing the required gate operations for Grover's algorithm, we
need to readout the computational results. This can be done by jointly
detecting the states of the three qubits \cite{Carlo,reed}. The readout for
flux qubit can be done by measuring the Josephson inductance of a SQUID that
is inductively coupled to the qubit \cite{lupas,lupas2,lupas3}. There are
some interesting schemes to perform joint dispersive readouts for
superconducting qubits \cite{blais, majer, filip}. The implementation of our
scheme also requires such joint dispersive readout for three SQUIDs.

\section{Imperfections in implementation}

Here, we discuss different types of imperfections which can arise during the
implementation of Grover's algorithm. The relevant parameters during the
implementation are coupling constant $g_{i},$ decay rate $\Gamma _{3}$ of
level $\left\vert 3\right\rangle $ and cavity decay rate $\kappa .$ We
consider spontaneous decay from the intermediate level $\left\vert
3\right\rangle $, during resonant interaction of cavity field with $%
\left\vert 2\right\rangle \leftrightarrow \left\vert 3\right\rangle $
transition of SQUIDs 1 and 2. Here, we assume that under the condition that
no photon from spontaneous emission is detected, the conditional Hamiltonian
for the evolution of system is given by \cite{yh}

\begin{equation}
H_{cond}=\hbar (g_{_{i}}a^{\dag }\left\vert 2\right\rangle _{i}\left\langle
3\right\vert +H.c,)-i\Gamma _{3}\left\vert 3\right\rangle _{i}\left\langle
3\right\vert .  \label{eq13}
\end{equation}

\begin{figure}[tbp]
\includegraphics[width=3.2 in]{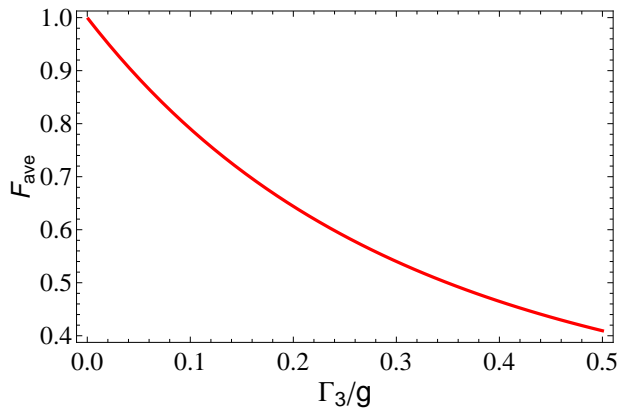}
\caption{Average fidelity $F_{ave}$\ of three-qubit controlled phase gate as
a function of coupling constant $\Gamma _{3}/g.$ }
\label{fig3}
\end{figure}

The Hamiltonian given by Eq. (\ref{eq13}) is valid as long as the decay rate
of level $\left\vert 2\right\rangle $ is much smaller than $\Gamma _{3}.$
The decay rate $\Gamma _{3}$ of SQUID 1 and 2 can affect the performance of
three-qubit phase gate. Suppose each SQUID is initially prepare in generic
state $\cos \nu \left\vert 0\right\rangle +\sin \nu \left\vert
1\right\rangle .$ Now the state of three qubits becomes $\left\vert \psi
(0)\right\rangle =a^{\prime }\left\vert 000\right\rangle +b^{\prime
}\left\vert 001\right\rangle +c^{\prime }\left\vert 010\right\rangle
+d^{\prime }\left\vert 011\right\rangle +e^{\prime }\left\vert
100\right\rangle +f^{\prime }\left\vert 101\right\rangle +g^{\prime
}\left\vert 110\right\rangle +h^{\prime }\left\vert 111\right\rangle ,$
where the coefficients $a^{\prime 3}\nu ,$ $b^{\prime }=c^{\prime
}=e^{\prime 2}\nu \sin \nu ,$ $d^{\prime }=f^{\prime }=g^{\prime 2}\nu $ and 
$h^{\prime 3}\nu $, satisfy the normalization condition. If we consider $%
\Gamma _{3}=0$, then the state of the system after phase gate operation
becomes $\left\vert \psi _{id}(\tau )\right\rangle =a^{\prime }\left\vert
000\right\rangle +b^{\prime }\left\vert 001\right\rangle +c^{\prime
}\left\vert 010\right\rangle +d^{\prime }\left\vert 011\right\rangle
+e^{\prime }\left\vert 100\right\rangle +f^{\prime }\left\vert
101\right\rangle +g^{\prime }\left\vert 110\right\rangle -h^{\prime
}\left\vert 111\right\rangle .$ However, if the decay of level $\left\vert
3\right\rangle $ is included during phase gate then the expression for $%
\left\vert \psi (\tau )\right\rangle $ becomes rather complex, therefore it
is not reproduced here. The fidelity of three-qubit phase gate is given by

\begin{eqnarray}
F &=&\left\vert \langle \psi _{id}(\tau )\left\vert \psi (\tau
)\right\rangle \right\vert ^{2}, \\
&=&\left\vert 1+\cos ^{2}\nu \sin ^{2}\nu (r^{4}-1)+\sin ^{4}\nu
(r^{2}-1)\right\vert ^{2}.
\end{eqnarray}%
Here, $r=(2g_{i}/\lambda )e^{-\eta }sin\theta $ with $\lambda =\sqrt{%
4g_{i}^{2}-\Gamma _{3}^{2}}$, $\eta =\pi \Gamma _{3}/4g_{i}$, and $%
\theta=\pi \lambda /4g_{i}.$ In order to realize the effects of decay on the
performance of three-qubit phase gate, average fidelity over all possible
three-qubit initial states is calculated using the following:

\begin{equation}
F_{ave}=\int\limits_{0}^{2\pi }d\varphi \int\limits_{0}^{\pi }\frac{F\sin
\nu d\nu }{4\pi }.
\end{equation}%
\bigskip After carrying out integration, we obtain

\begin{equation}
F_{ave}=\frac{1}{315}(63+48r^{2}+164r^{4}+32r^{6}+8r^{8}).
\end{equation}%
Next, we show the plots of average fidelity as a function of $\Gamma _{3}/g$
in Fig. \ref{fig3}. It can easily be verified that for $\Gamma _{3}=0,$ one
has $r=1,$ which leads to $F=F_{ave}=1.$ It is clear from Fig. \ref{fig3}
that average fidelity decreases due to the increase in the cavity decay
rate. We also calculate the success probability of three-qubit phase gate
which is given by

\begin{eqnarray}
P &=&(\left\vert a^{\prime }\right\vert ^{2}+\left\vert b^{\prime
}\right\vert ^{2}+\left\vert c^{\prime }\right\vert ^{2}+\left\vert
d^{\prime }\right\vert ^{2}+r^{8}\left\vert e^{\prime }\right\vert ^{2} 
\notag \\
&&+r^{8}\left\vert f^{\prime }\right\vert ^{2}+r^{4}\left\vert g^{\prime
}\right\vert ^{2}+r^{4}\left\vert h^{\prime }\right\vert ^{2}).  \label{eq14}
\end{eqnarray}%
If we consider $\nu =\pi /4,$ then corresponding success probability of
three-qubit phase gate reduces to $P=(4+2r^{4}(r^{4}+1))/8.$

\begin{figure}[tbp]
\includegraphics[width=3.6 in]{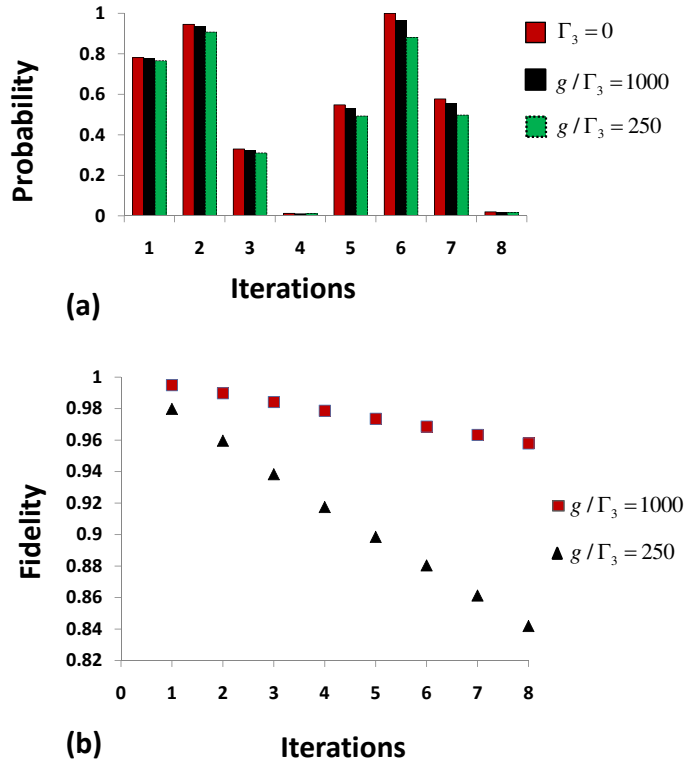}
\caption{Simulations results for three-qubit Grover's algorithm involving
spontaneous decay rate $\Gamma _{3}$\ from intermediate level $|3\rangle .$
(a) Probability of finding the target state $|111\rangle $\ in case of $%
\Gamma _{3}/g=0$, $0.001$, and $0.004.$ (b) Fidelity of the state searched
for $\Gamma _{3}/g=0.001$ and $0.004.$}
\label{fig4}
\end{figure}

We take into account success rate of three-qubit phase gate (Eq. \ref{eq14})
and Grover search itself. We consider $g_{1}\thicksim g_{2}\thicksim g,$
without loss of generality and perform simulation for finding the target
state for $\Gamma _{3}/g=0$ (ideal case), $0.001$, and $0.004.$ Here, we
consider the success probability for target state $\left\vert
111\right\rangle $ as shown in Fig. \ref{fig4}(a). It is clear from Fig. \ref%
{fig4}(a) that success rate becomes closer to the ideal case when decay rate
is sufficiently smaller than coupling constant i.e., $\Gamma _{3}/g=0.001.$
Typical decoherence rate for SQUID is of the order $10^{6}s^{-1}$ while
coupling constant can be achieved upto $3\times 10^{9}s^{-1}$ \cite{han}.
This shows that for these parameters, three-qubit controlled phase gate and
Grover's iterations can be performed with high fidelity. Probability of
success is highest at $6th$ iteration for ideal case. However, we should
prefer $2nd$ iteration in the presence of dissipation because it has the
highest value of fidelity. The effect of level decay on the fidelity of
search state during iteration is shown in Fig. \ref{fig4}(b). It is clear
from Fig. \ref{fig4}(b) that the fidelity of state to be searched decreases
much rapidly for the case of larger decay rate as compared to smaller decay
rate as the number of iterations increase.

\begin{figure}[tbp]
\includegraphics[width=3.2 in]{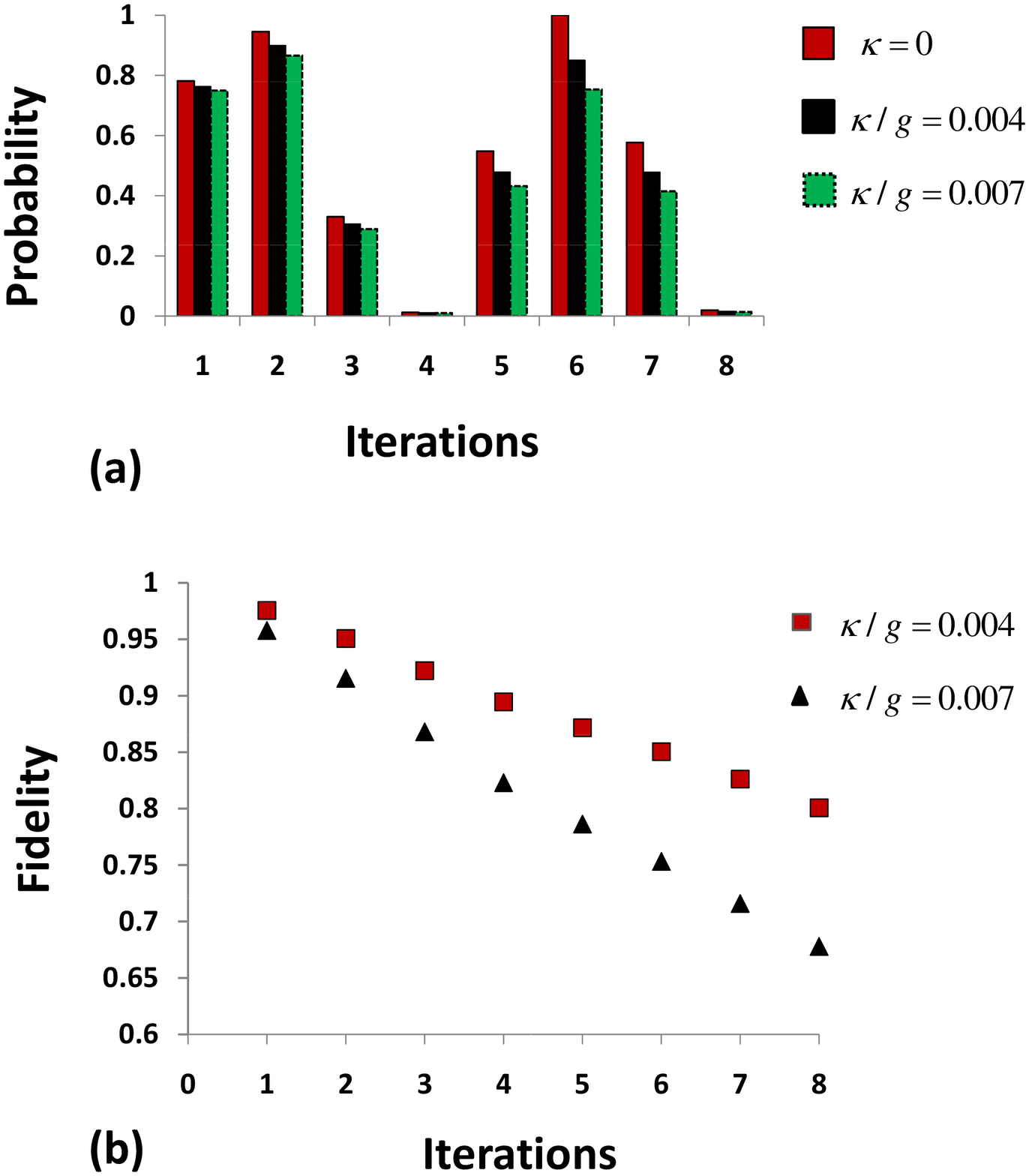}
\caption{Simulation results for three-qubit Grover's algorithm involving
cavity decay rate $\protect\kappa $ (a) Probability of finding the target
state $|111\rangle $\ in case of $\protect\kappa /g=0$, $0.004,$ and $0.007.$
(b) Fidelity of the state searched for $\protect\kappa /g=0.004$ and $0.007.$%
}
\label{fig5}
\end{figure}

Next we consider the effects of cavity decay during the implementation of
Grover's algorithm. During the implementation of three-qubit phase gate
(Sec. IV), transition of SQUID 1 from level $\left\vert 3\right\rangle $ to $%
\left\vert 2\right\rangle $ would result in the emission of one photon (See
step $1$ in Sec. IV). Then transition of SQUID 2 from level $\left\vert
2\right\rangle $ to $\left\vert 3\right\rangle $ would absorb this photon
with unit probability in step $3$ and vice versa for step $6$ and $8.$ In
the absence of cavity decay the occupation probability of level $\left\vert
2\right\rangle $ and $\left\vert 3\right\rangle $ of SQUID 1 and 2 should be
exactly one. However, if cavity relaxation is taken into account, then the
occupation probabilities are expected to decay exponentially. Under the
assumption that no photon actually leaks out during implementation time, we
can write the conditional Hamiltonian as \cite{wlyang}

\begin{equation}
H_{c}=\hbar (g_{_{i}}a^{\dag }\left\vert 2\right\rangle _{i}\left\langle
3\right\vert +H.c,)-i\kappa a^{\dag }a,  \label{ch}
\end{equation}%
where, $\kappa $ is the cavity decay rate and $g_{i}$ is the coupling
constant. For $g_{i}\gg \kappa ,$ we only need to focus on time evolution of
the system governed by conditional Hamiltonian (Eq. (\ref{ch})) under the
assumption of strong coupling limit. The fidelity and corresponding success
probabilities for three-qubit phase gate can easily be obtained, which are
given by

\begin{equation}
F_{ave}=\frac{(4+2e^{-2\kappa t}+e^{-\kappa t}+e^{-3\kappa t/2})^{2}}{%
8(4+2e^{-4\kappa t}+e^{-2\kappa t}+e^{-3\kappa t})},
\end{equation}
and

\begin{equation}
P=\frac{1}{8}(4+2e^{-4\kappa t}+e^{-2\kappa t}+e^{-3\kappa t}).
\end{equation}%
For $\kappa =0$, we have $F_{ave}=P=1.$ The success probability of target
state $\left\vert 111\right\rangle $ for ideal case and in the presence of
cavity decay is shown in Fig. \ref{fig5}(a). It can be seen that the
probability of finding target state decreases in the presence of cavity
decay rate. The effects of cavity decay rate on fidelity of Grover's search
iteration is also shown in Fig. \ref{fig5}(b). It is clear from Fig. \ref%
{fig5}(b) that the fidelity of Grover's search decreases as a function of
iterations much rapidly for higher cavity decay rate.

\begin{figure}[tbp]
\includegraphics[width=3.2 in]{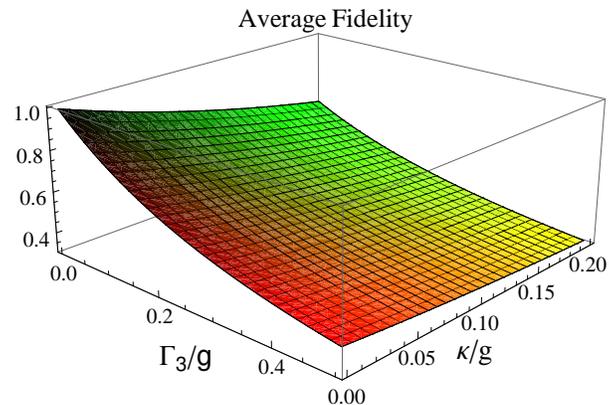}
\caption{Average fidelity $F_{ave}$\ of three-qubit controlled phase gate as
a function of cavity and level decay}
\label{fig6}
\end{figure}

We have separately discussed the effects on fidelity from spontaneous decay
and cavity decay, however, in a general case the system involves both these
decays. Thus we also consider the effect of these decays simultaneously. The
results of our numerical simulation for corresponding average fidelity are
shown in Fig. \ref{fig6}. It is clear from Figs. \ref{fig4}-\ref{fig6} that
our proposed Grover's iterations can be performed with high success
probability and fidelity as long as cavity decay rate and decay rate of the
intermediate level $\left\vert 3\right\rangle $ is small enough. The typical
values of cavity decay rate is $\kappa ^{-1}\thicksim 0.76\mu s$ ($%
Q\thicksim 10^{5}$) \cite{han}. It may be mentioned that more rigorous
analysis is required for the case of very low $Q$ resonators.

\section{DISCUSSION AND CONCLUSION}

We have discussed the implementation of 3-qubit Grover's algorithm using
4-level SQUIDs subjected to quantized and classical microwave fields.
Grover's algorithm involves three-qubit phase gates and single-qubit gates.
Here, we briefly estimate the total operational time for three-qubit
controlled phase gate. The total implementation time is the sum of all
interaction times involved in three-qubit controlled phase gate operation
i.e., $\tau =2t_{1}+2t_{1}^{^{\prime }}+2t_{2}+2t_{3}^{^{\prime
}}+2t_{3}+2t_{4}+t_{4}^{\prime }.$ On substituting the values of interaction
times given in Sec. $IV$ we obtain 
\begin{equation}
\tau =\pi (\frac{1}{\Omega _{13}}+\frac{1}{g_{1}}+\frac{1}{\Omega _{02}}+%
\frac{1}{g_{2}}+\frac{1}{\Omega _{30}}+\frac{1}{\Omega _{12}}+\frac{\Delta
_{c}}{g_{3}^{2}}).
\end{equation}%
Here, we consider without loss of generality $g_{1}\thicksim g_{2}\thicksim
g_{3}\thicksim 3\times 10^{9}s^{-1}$ which is the same as given in Ref \cite%
{han}. Choosing $\Delta _{c}=10g_{3}$, $\Omega _{02}\thicksim \Omega
_{13}\thicksim \Omega _{12}\thicksim \Omega _{30}\thicksim 10g_{1}$, the
operational time for three-qubit phase gate comes out to be $\tau \thicksim
13ns$. The operation time for single-qubit gate is about $1.5ns$ \cite{yh}
which can be applied to each qubit, simultaneously. Therefore, the estimated
time for the implementation of three-qubit Grover algorithm performing two
iterations is approximately $66ns$.

We have considered the effect of decay for level $\left\vert 3\right\rangle$
under the assumption that the decay rate of level $\left\vert 2\right\rangle$
is much smaller then level $\left\vert 3\right\rangle$. The typical values
of the decay time for levels $\left\vert 3\right\rangle$ and $\left\vert
2\right\rangle$ are $\Gamma_{3}^{-1}\thicksim 3.2\mu s$ and $\Gamma
_{2}^{-1}\thicksim$ 0.16 ms as discussed in Ref. \cite{han}. During the
steps 1,2, and 3 in phase gate operations level $\left\vert 2\right\rangle$
of SQUIDs 1 and 2 is occupied through the application of SQUID-pulse
resonant interaction and SQUID-resonator resonant interaction as discussed
in Sec. IV. Operation time $t_{2}+t_{1}^{^{\prime}}$ for SQUID 1 and $%
t_{2}+t_{3}^{^{\prime}}$ for SQUID 2, in these steps is equal to $\pi
/(2\Omega _{20})+\pi /(2g_{1})$ and $\pi /(2\Omega _{20})+\pi /(2g_{2})$,
respectively, which can be shortened by increasing the Rabi frequency $%
\Omega _{20}$ and coupling constants $g_{1}$ and $g_{2}$. For the typical
choice of parameters as given in Ref \cite{han}, we have $%
t_{2}+t_{1}^{^{\prime }}\thicksim t_{2}+t_{3}^{^{\prime }}\thicksim 0.6ns$,
which is much shorter than $\Gamma _{2}^{-1}$. The SQUIDs can also be
designed to have long relaxation time for level $\left\vert 2\right\rangle$.
Thus decoherence due to relaxation of level $\left\vert 2\right\rangle $ can
be negligibly small. As far as the decay of level $\left\vert 1\right\rangle$
is concerned, it may be pointed out that in our scheme direct coupling
between levels $\left\vert1\right\rangle$ and $\left\vert 0\right\rangle$ is
not needed. The potential barrier between levels $\left\vert 1\right\rangle$
and $\left\vert 0\right\rangle$ can also be adjusted such that decay of
level $\left\vert 1\right\rangle$ is negligibly small \cite{lap,han}.
Therefore, storage time of each qubit can be made much longer.

When levels $\left\vert 2\right\rangle $ and $\left\vert 3\right\rangle $
are manipulated by microwave pulses, resonant interaction between cavity
field mode and $\left\vert 2\right\rangle \leftrightarrow \left\vert
3\right\rangle $ transition of each control SQUID is unwanted. This effect
can be minimized by setting the condition $\Omega _{i,j}>>g$ for SQUIDs 1
and 2.

\begin{figure}[tbp]
\includegraphics[width=3.2 in]{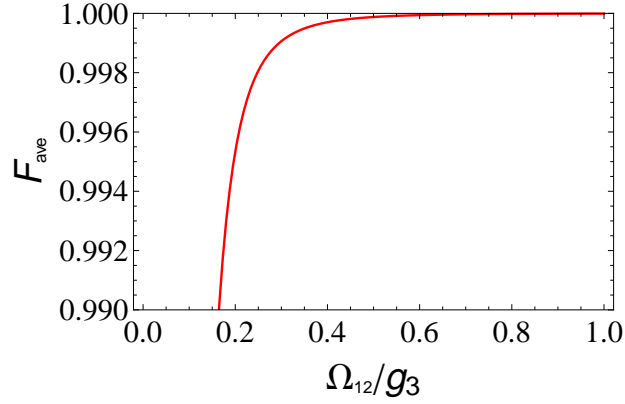}
\caption{Average fidelity $F_{ave}$\ of three-qubit controlled phase gate as
function of Rabi frequency $\Omega _{12}$\ for the case of $\Delta
_{c}=10g_{3}.$}
\label{fig7}
\end{figure}

During the application of microwave pulse, off-resonant interaction of
cavity field with $\left\vert 2\right\rangle \leftrightarrow \left\vert
3\right\rangle $ transition, in the presence of single photon inside the
cavity is unwanted. It induces an unwanted phase which can effect the
performance of the desired gate. This effect can be negligible under the
condition $\Omega _{12}>>g_{3}^{2}/\Delta _{c}$ for SQUID $3$. However, if
this effect is included during the steps $4$ and $5$ for three-qubit phase
gate, then the corresponding fidelity can easily be obtained. Here, we do
not present the expression for $\left\vert \psi (\tau )\right\rangle$ owing
to its complexity, however the fidelity is given by

\begin{equation}
F(x)=|1-x-x(\alpha ^{2}-\beta ^{2}-\gamma ^{2}))|^{2},
\end{equation}
where, $x=|h|^{2},$ $\alpha =\cos \xi ,$ $\beta =\delta \sin \xi /\sqrt{%
\delta ^{2}+4\Omega _{12}^{2}}$, $\gamma =2\Omega _{12}\sin \xi /\sqrt{%
\delta ^{2}+4\Omega _{12}^{2}}$ with $\delta =g_{3}^{2}/\Delta _{c}$ and $%
\xi =\pi \sqrt{\delta ^{2}+4\Omega _{12}^{2}}/4\Omega _{12}.$ The average
fidelity over all possible initial states is given by

\begin{eqnarray}
F_{ave} &=&\int\limits_{0}^{1}F(x)dx=\frac{1}{3}(1+\alpha ^{4}+\beta
^{4}+\gamma ^{2}+\gamma ^{4}  \notag \\
&&+\beta ^{2}(1+2\gamma ^{2})-\alpha ^{2}(1+2\beta ^{2}+2\gamma ^{2})).
\end{eqnarray}%
If off-resonant interaction during step $4$ and $5$ is not considered, then
we have $\delta =0,$ $\alpha =\beta =0,$ and$\ \gamma =1$ which leads to $%
F=F_{ave}=1.$ The plot of the average fidelity as a function of Rabi
frequency $\Omega _{12}/g_3$ is shown in Fig. \ref{fig7}. We choose $\Delta
_{c}=10g_{3}$ for this plot. It can be seen that the average fidelity
increases as a function of Rabi frequency $\Omega _{12}$ applied to target
SQUID. $\ $It is clear from Fig. \ref{fig7} that for $\Omega _{12}=0.6g_{3}$%
, we have $F_{ave}\thicksim 1.$

When slowly changing Rabi frequencies are applied to satisfy the adiabatic
passage, gate times becomes slow i.e., of the order of 1 ms to a few
microseconds \cite{wang}. However, in our scheme we do not require slowly
changing Rabi frequency during the implementation of three-qubit phase gate,
thus gate is significantly faster i.e., of the order of 13ns. The fast
pulses may introduce new imperfections, for example, accurately designing of
the duration of pulses might not be easy. It may be mentioned here that
adiabatic process is not always slow as an interesting proposal based on
controllable Stark-chirped rapid adiabatic passage has been proposed in a
recent study \cite{lfw}.

Here, we would like to mention that the physical implementation of
three-qubit Grover's algorithm in cavity quantum electrodynamics (QED) has
been proposed in a recent study by Yang \textit{et al} \cite{wlyang}. The
scheme is based upon the resonant interaction of three Rydberg atoms
initially prepared in a coherent superposition state traversing through a
single-mode microwave cavity. As compared to the flying qubits, here we have
considered stationary qubits defined through two lowest level of four level
SQUIDs. Our scheme is based on the generation and absorption of single
photon in high Q cavity using SQUIDs. The generation of single microwave
photon in superconducting qubit have been reported, experimentally in some
recent studies \textit{et al}. \cite{photon,photon2}.

In conclusion, we have proposed a scheme for the realization of three-qubit
Grover's algorithm based on three-qubit phase gate using four-level SQUIDs
coupled to a single-mode superconducting resonator. In this proposal,
adjustment of level spacing during the operation, slowly changing Rabi
frequencies (to satisfy adiabatic passage), and the use of second-order
detuning (to achieve off-resonance Raman coupling between two relevant
levels) are not required. Thus, implementation time is significantly faster
and has operation time of the order of nanoseconds. The coupling constants
of each SQUID with the resonator are different. Thus, an unavoidable
non-uniformity in device parameters can be accommodated. We consider the
effect of imperfections in the system which include the decay of the cavity
field and the relevant level. We also incorporate the influence of unwanted
off-resonant interaction during the gate implementation. Our results show
that the marked state can be searched with high fidelity even in the
presence of imperfections in the system which is quite interesting.

\end{document}